\documentclass[11pt]{article}
\usepackage[dvips]{color}
\usepackage{epsfig}
\usepackage{amsmath}
\usepackage{graphicx}

\begin{document}
\title{Particle Acceleration in Rotating Modified Hayward and Bardeen Black Holes}
\author{{Behnam Pourhassan$^{a}$\thanks{Email:
b.pourhassan@du.ac.ir}\hspace{1mm} and Ujjal
Debnath$^{b}$\thanks{Email: ujjaldebnath@gmail.com ;
ujjal@iucaa.ernet.in}}\\
$^{a}${\small {\em  School of Physics, Damghan University, Damghan, 3671641167,  Iran.}}\\
$^{b}${\small {\em Department of Mathematics, Indian Institute of Engineering Science and Technology,}}\\
{\small {\em Shibpur, Howrah-711 103, India.}}} \maketitle

\begin{abstract}
In this paper we consider rotating modified Hayward and rotating
modified Bardeen black holes as particle accelerators. We
investigate the center of mass energy of two colliding neutral
particles with same rest masses falling from rest at infinity to
near the horizons of the mentioned black holes. We investigate the
range of the particle's angular momentum and the orbit of the
particle. We also investigate
the center of mass energy for extremal black hole.\\\\
\noindent {Keywords:} Black hole; Particle acceleration.
\end{abstract}

\section{Introduction}
Collision of particles around black holes is very interesting
topic in current astrophysical research. Recently,
Ba$\tilde{n}$ados, Silk and West (BSW) \cite{BSW} have
investigated the collision of two particles falling from rest at
infinity into the Kerr black hole, which is known as BSW
mechanism. They determined the center of mass (CM) energy in the
equatorial plane, which may be high in the limiting case of
extremal black hole. Further Lake \cite{Lake1,Lake2} demonstrated
that the CM energy of two colliding particles diverges at the
inner horizon of non-extremal Kerr black hole. A general review of
particle accelerator of black hole is discussed by Harada et al
\cite{Harada}. Wei et al \cite{Wei} also investigated that the
collision of two uncharged particles around Kerr-Newmann black
hole, which properly depends on the spin and charge of the black
hole. Liu et al \cite{Liu} demonstrated that the collision of two
particles around Kerr-Taub-NUT black hole. Subsequently, Zakria
and Jamil \cite{Zak} investigated the CM energy of the collision
for two neutral particles with different rest masses falling
freely from rest at infinity in the background of a
Kerr-Newman-Taub-NUT black hole. Till now, several authors
\cite{Berti, Jacob, Zas, Igata, Ban, Hus, Grib, Shar, Wei1, Ghosh,
Pour1, Pour2} have studied the collision of particles near black
holes and CM energy for the colliding particles.\\

It was argued that the CM energy of the colliding particles for
the naked singularity diverges \cite{P1,P2,P3,P4}. Due to
gravitational collapse, any astrophysical object produces either
space-time singularity or naked singularity. Since we know that
any classical black hole has a singularity. To avoid the
singularity, Bardeen \cite{Bardeen} proposed the concept of
regular black hole, dubbed as Bardeen black hole and subsequently,
another type of regular black hole (Hayward black hole) found
\cite{Hayward}. Another kind of regular black hole is Ayon-Beato-
Garcya (ABG) black hole \cite{Ayon}. Geodesic study of regular
Hayward black hole has been discussed by Abhas et al \cite{Abhas}.
The implication of rotating Hayward black hole is discussed in
ref.\cite{Bambi}. Modified Hayward black hole metric has been
proposed by Lorenzo et al \cite{Lor}. Recently, Amir et al
\cite{Amir} studied the collision of two particles with equal
masses moving in the equatorial plane near horizon of the rotating
Hayward's regular black hole (as particle accelerator). Also P.
Pradhan \cite{Pradhan} studied the regular Hayward and Bardeen
black holes as particle accelerator. The CM energy of the
collision for charged particles in a Bardeen black hole was
studied in \cite{Hai}. CM energy and horizon structure for
rotating Bardeen black hole are also studied in ref \cite{G}. We
now extend the above works into rotating modified Hayward and
rotating modified Bardeen black holes. The CM energy and the
particles orbit are investigated for two colliding neutral
particles of same rest masses falling from infinity into the above
mentioned black holes. Next we discuss the extremal limits of the
black holes. Finally we conclude the results for particle
acceleration near the black holes.

\section{Rotating Black Hole Background}
The rotating black hole metric can be written as
\begin{equation}\label{R1}
ds^{2}=g_{tt}dt^{2}+g_{rr}dr^{2}+g_{\theta\theta}d\theta^{2}+g_{\phi\phi}d\phi^{2}+2g_{t\phi}dtd\phi,
\end{equation}
where,
\begin{eqnarray}\label{R2}
g_{tt}&=&-\tilde{h}\tilde{f},\nonumber\\
g_{rr}&=&\frac{\Sigma}{\tilde{f}\Sigma+a^{2}\sin^{2}\theta},\nonumber\\
g_{\theta\theta}&=&\Sigma,\nonumber\\
g_{\phi\phi}&=&\sin^{2}\theta\left[\Sigma+a^{2}(2-\tilde{f})\sin^{2}\theta\right],\nonumber\\
g_{t\phi}&=&a(1-\tilde{h}\tilde{f})\sin^{2}\theta,
\end{eqnarray}
with
\begin{equation}\label{R3}
\Sigma=r^{2}+a^{2}\cos^{2}\theta,
\end{equation}
and
\begin{eqnarray}\label{R4}
\tilde{f}&=&1-\frac{2\tilde{m}_{1}r}{\Sigma},\nonumber\\
\tilde{h}&=&1-\frac{\mu\tilde{m}_{2}r}{\Sigma^{2}}.
\end{eqnarray}
In this paper we are interested to investigate two different but
approximately similar black holes: rotating modified Hayward and
Bardeen black holes where $\tilde{m}_{1}$ and $\tilde{m}_{2}$ are
different for them which are explained below.

\subsection{Rotating Modified Hayward Black Hole}

The rotating modified Hayward black hole metric is defined in
equation (\ref{R1}) provided $\tilde{m}_{1}$ and $\tilde{m}_{2}$
are given by \cite{Bambi, Amir}
\begin{eqnarray}\label{R5}
\tilde{m}_{1}&=&M\frac{r^{3+\alpha}\Sigma^{-\frac{\alpha}{2}}}{r^{3+\alpha}
\Sigma^{-\frac{\alpha}{2}}+g_{1}^{3}r^{\beta}\Sigma^{-\frac{\beta}{2}}},\nonumber\\
\tilde{m}_{2}&=&M\frac{r^{3+\alpha}\Sigma^{-\frac{\alpha}{2}}}{r^{3+\alpha}
\Sigma^{-\frac{\alpha}{2}}+g_{2}^{3}r^{\beta}\Sigma^{-\frac{\beta}{2}}}
\end{eqnarray}
Here $g_{1}^{3}=2Ml^{2}$ and $g_{2}^{3}=\frac{\mu}{\nu}M$, where
$\mu$ and $\nu$ are positive constants, and $l$ is a parameter
with dimensions of length with small scale related to the inverse
cosmological constant and also $\alpha$ and $\beta$ are real
numbers. It is easy to check that $\alpha=\beta=a=0$ yield to
non-rotating Hayward black hole and $\mu=0$ reduced to ordinary
Hayward black hole. Horizon structure of rotating modified Hayward
black hole given by $g_{rr}=\infty$ from (\ref{R2}) which is
exactly similar to the rotating Hayward black hole discussed by
the Ref \cite{Amir}.

\subsection{Rotating Modified Bardeen Black Hole}

\begin{figure}[h!]
 \begin{center}$
 \begin{array}{cccc}
\includegraphics[width=50 mm]{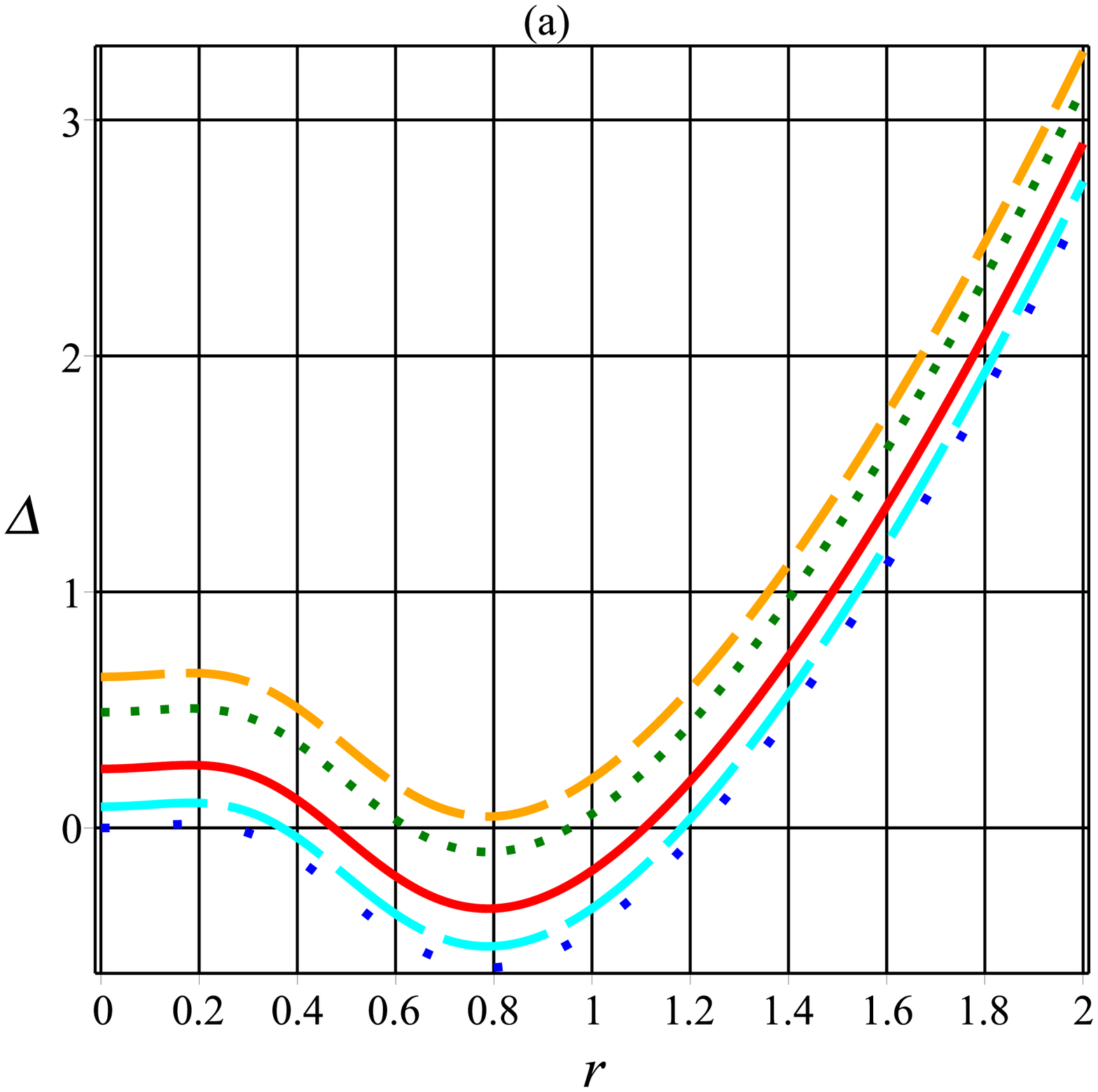}&\includegraphics[width=50 mm]{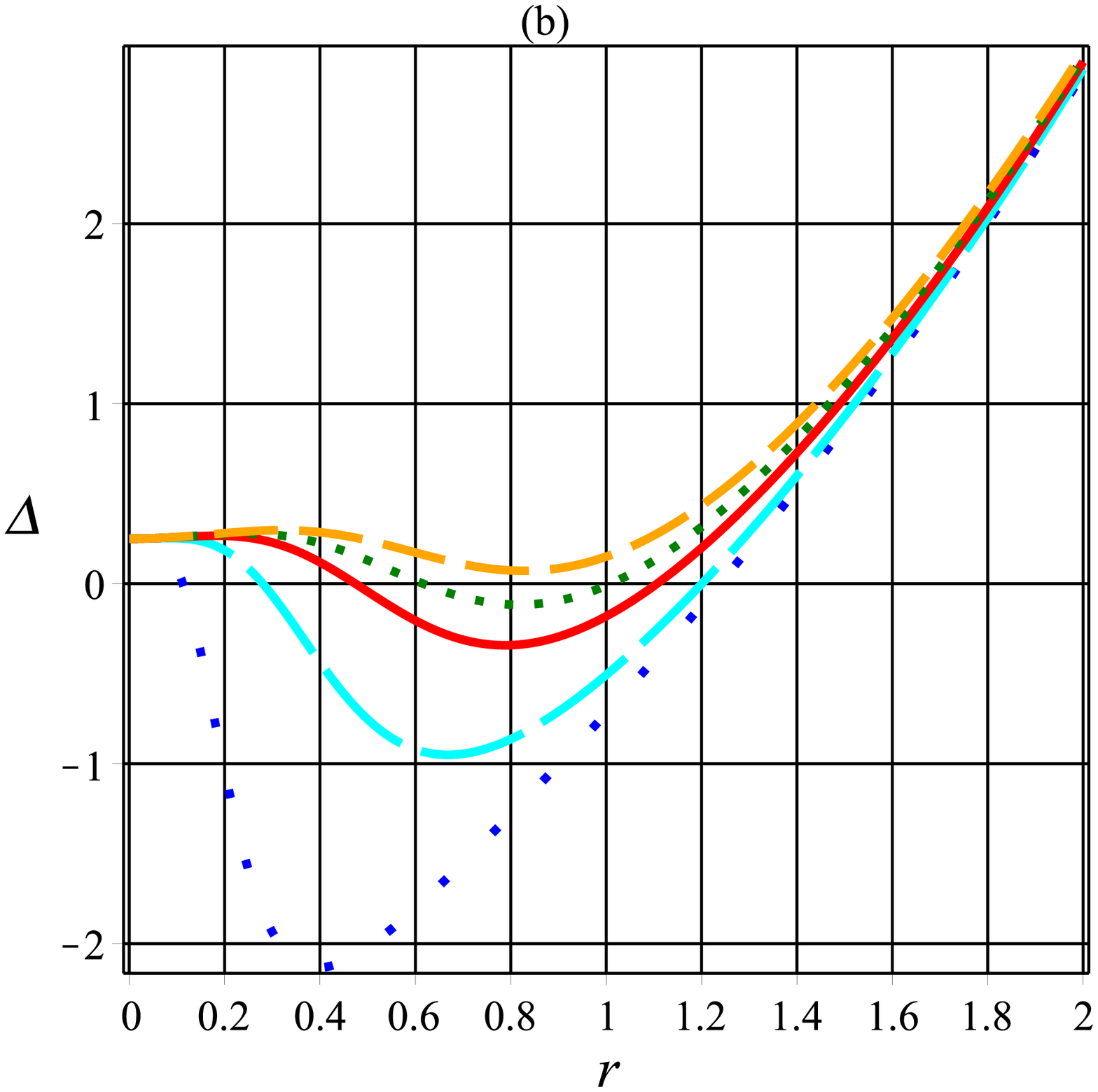}\\
\includegraphics[width=50 mm]{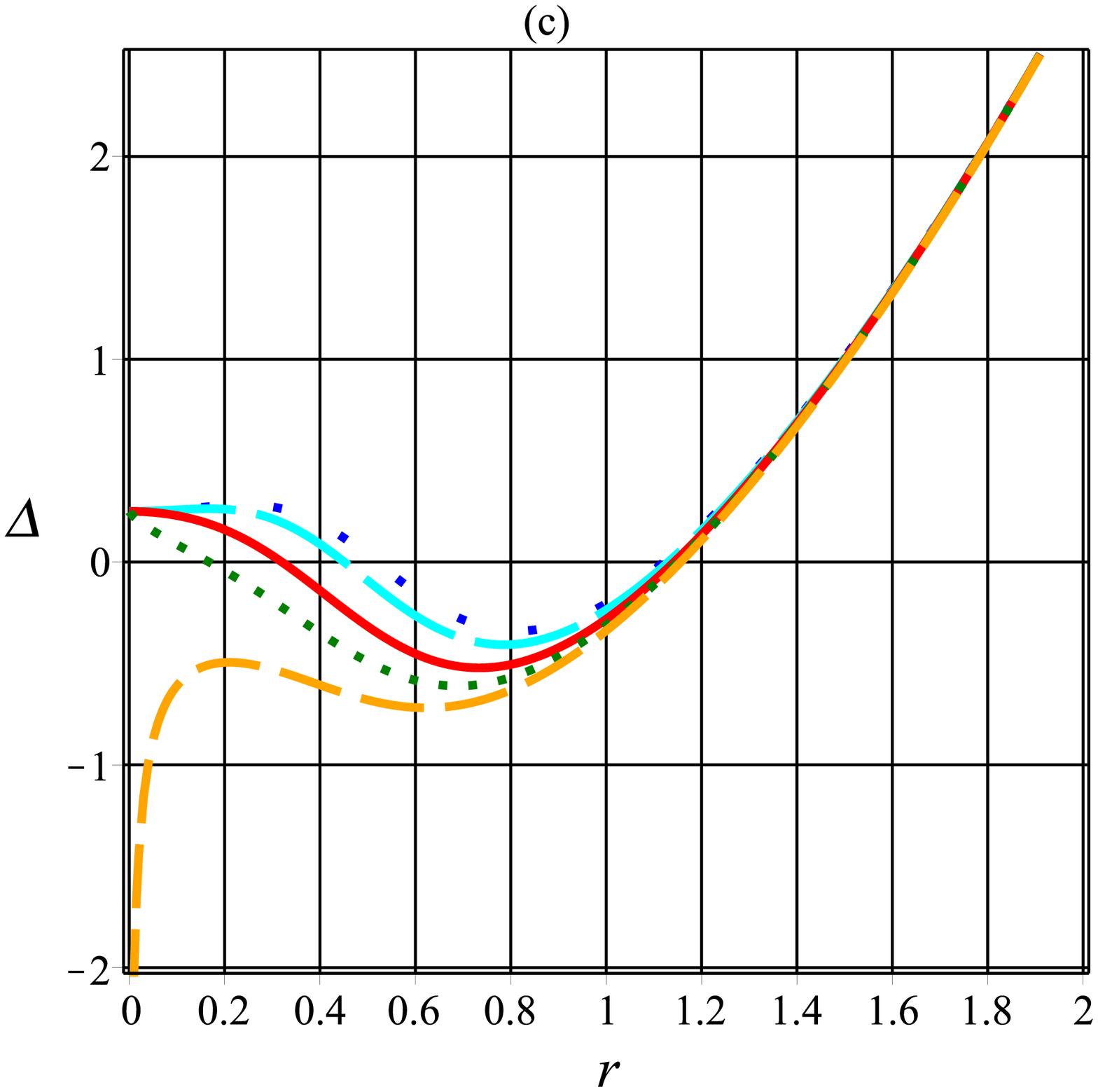}&\includegraphics[width=50 mm]{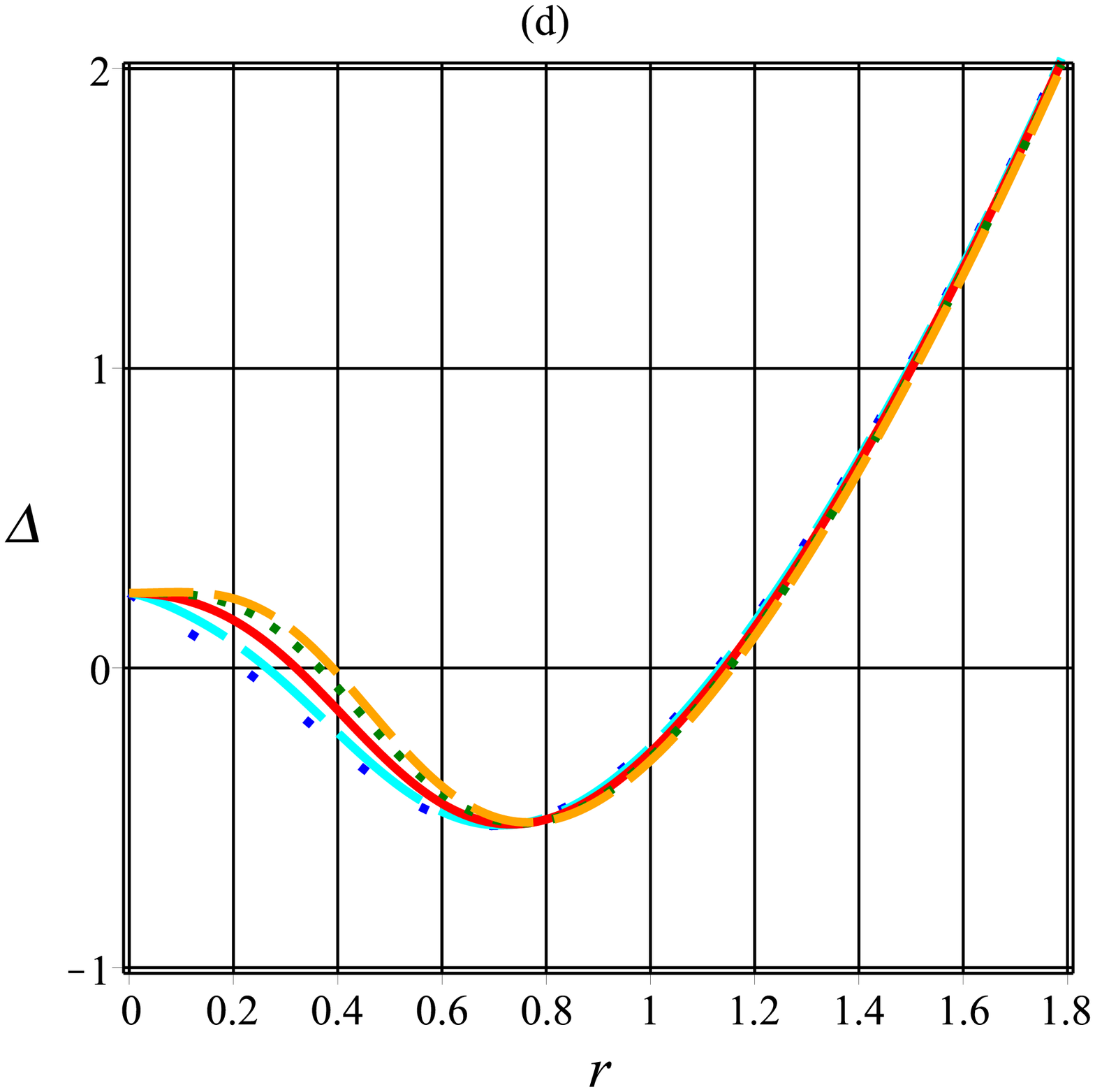}
 \end{array}$
 \end{center}
\caption{$\Delta$ in terms of $r$ for various values of $l$, $a$, $\alpha$ and
$\beta$ with $M=1$, $\theta=\frac{\pi}{2}$ ((a) and (b)), $\theta=\frac{\pi}{6}$
((c) and (d)). (a) $\alpha=1$ and $\beta=2$, $l=0.5$, $a=0$ (blue space dot),
$a=0.3$ (cyan long dash), $a=0.5$ (red solid), $a=0.7$ (green dot), $a=0.8$
(orange dash). (b) $\alpha=1$ and $\beta=2$, $a=0.5$, $l=0.1$ (blue space dot),
$l=0.3$ (cyan long dash), $l=0.5$ (red solid), $l=0.6$ (green dot), $l=0.7$
(orange dash). (c) $a=0.5$, $l=0.5$, $\alpha=1$, $\beta=0$ (blue space dot),
$\beta=0.8$ (cyan long dash), $\beta=2$ (red solid), $\beta=2.8$ (green dot),
$\beta=3.6$ (orange dash). (d) $a=0.5$, $l=0.5$, $\beta=2$, $\alpha=0$ (blue space dot),
$\alpha=0.4$ (cyan long dash), $\alpha=1$ (red solid), $\alpha=1.6$ (green dot),
$\alpha=2$ (orange dash).}
 \label{fig:1}
\end{figure}

The rotating modified Bardeen black hole metric is defined in
equation (\ref{R1}) provided $\tilde{m}_{1}$ and $\tilde{m}_{2}$
are given by
\begin{eqnarray}\label{R6}
\tilde{m}_{1}&=&M\frac{r^{3+\alpha}\Sigma^{-\frac{\alpha}{2}}}
{\left(r^{3+\alpha}\Sigma^{-\frac{\alpha}{2}}+Ml^{2}r^{\beta}
\Sigma^{-\frac{\beta}{2}}\right)^{\frac{3}{2}}},\nonumber\\
\tilde{m}_{2}&=&M\frac{r^{3+\alpha}\Sigma^{-\frac{\alpha}{2}}}
{\left(r^{3+\alpha}\Sigma^{-\frac{\alpha}{2}}+g^{2}r^{\beta}
\Sigma^{-\frac{\beta}{2}}\right)^{\frac{3}{2}}}
\end{eqnarray}
where, as before, $l$ is a parameter with dimensions of length,
$\alpha$ and $\beta$ are real numbers, and $g$ is a constant
parameter with mass dimension. Horizons of the black hole given by
root of the following equation (denominator of $g_{rr}$ to be
zero),
\begin{equation}\label{R7}
\tilde{f}\Sigma+a^{2}\sin^{2}\theta=0.
\end{equation}
Using the equation (\ref{R6}) we have
\begin{equation}\label{R8}
\Delta=r^{2}+a^{2}-\frac{2Mr^{4+\alpha}\Sigma^{-\frac{\alpha}{2}}}
{\left(r^{3+\alpha}\Sigma^{-\frac{\alpha}{2}}+Ml^{2}r^{\beta}
\Sigma^{-\frac{\beta}{2}}\right)^{\frac{3}{2}}}=0.
\end{equation}
Horizon structure of rotating modified Bardeen black hole given by
the plots of Fig. \ref{fig:1}. We can see that for the suitable
choice of parameters there are two horizons can be written as
$r_{\pm}=r\pm \delta$, where $0<\delta<0.5$. The case of
$\theta=\frac{\pi}{2}$ is our interest, although values of
$\alpha$ and $\beta$ are not important. In this case, $\Delta$ vs
$r$ is plotted in Figs. \ref{fig:1} (a) and (b). On the other hand
for the case of $\theta=\frac{\pi}{6}$ we can see effect of
$\alpha$ and $\beta$ on black hole horizons in Figs. \ref{fig:1}
(c) and (d). For example from red solid lines of Fig. \ref{fig:1} (a) and (b) we can see $r_{+}\approx1.1$ and $r_{-}\approx0.5$ for $\alpha=1$, $\beta=2$ and $a=0.5$ (Bardeen black hole). Also one can obtain $r_{+}\approx1.65$ and $r_{-}\approx0.9$ with $\alpha=1$, $\beta=2$ and $a=0.5$ for the case of Hayward black hole.

\section{The Center of Mass Energy}
\begin{figure}[h!]
 \begin{center}$
 \begin{array}{cccc}
\includegraphics[width=50 mm]{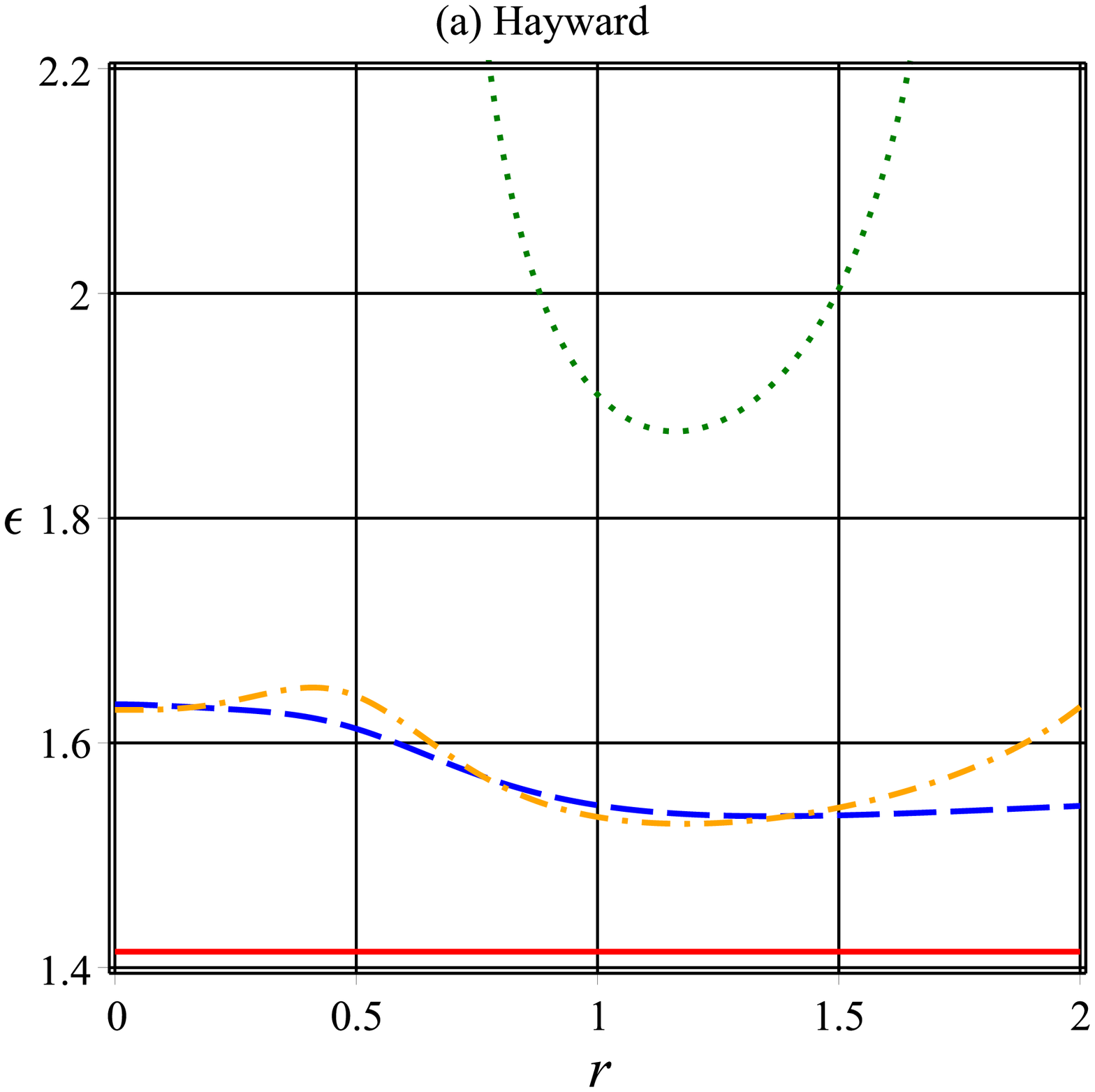}&\includegraphics[width=50 mm]{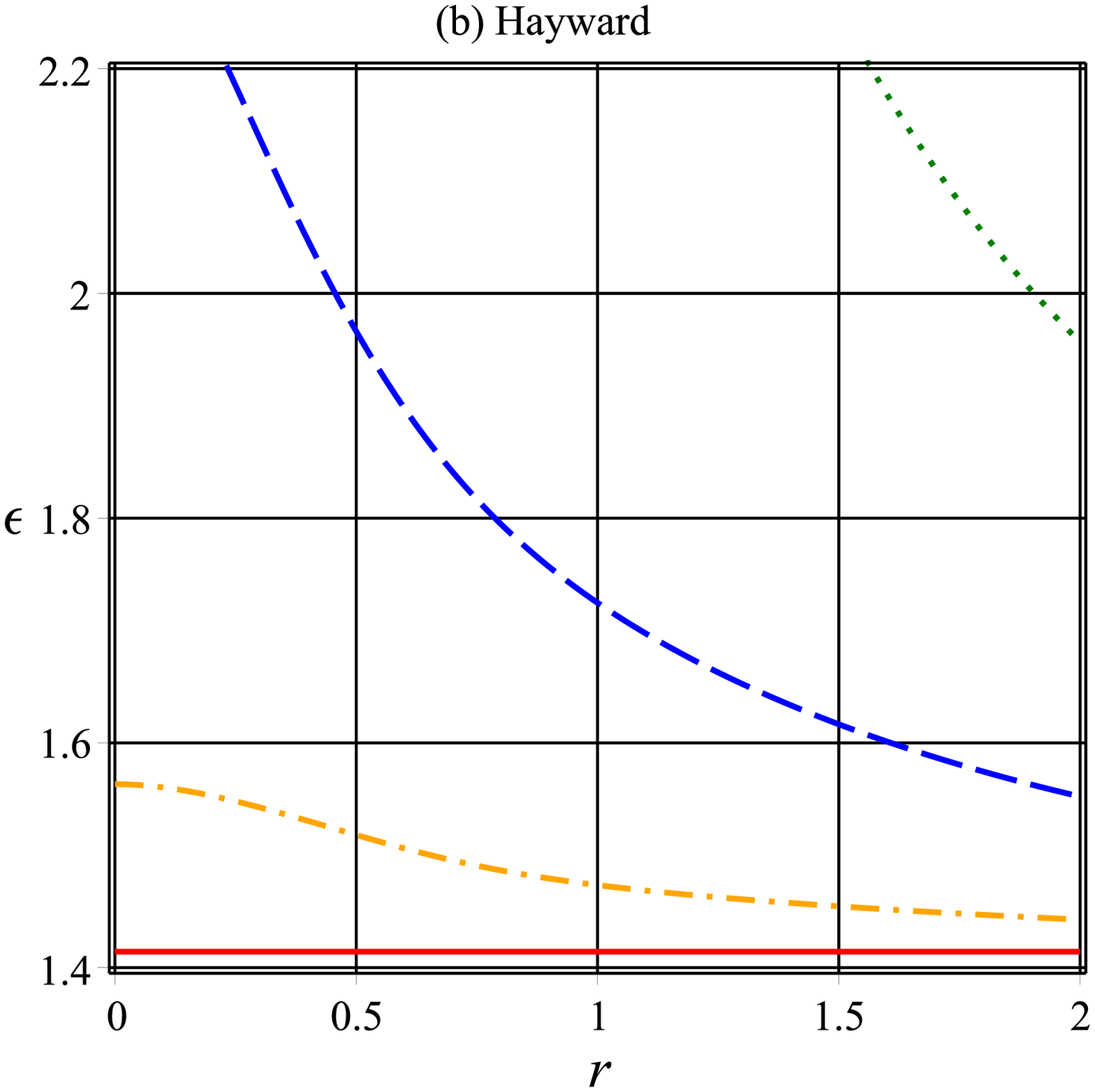}&\includegraphics[width=50 mm]{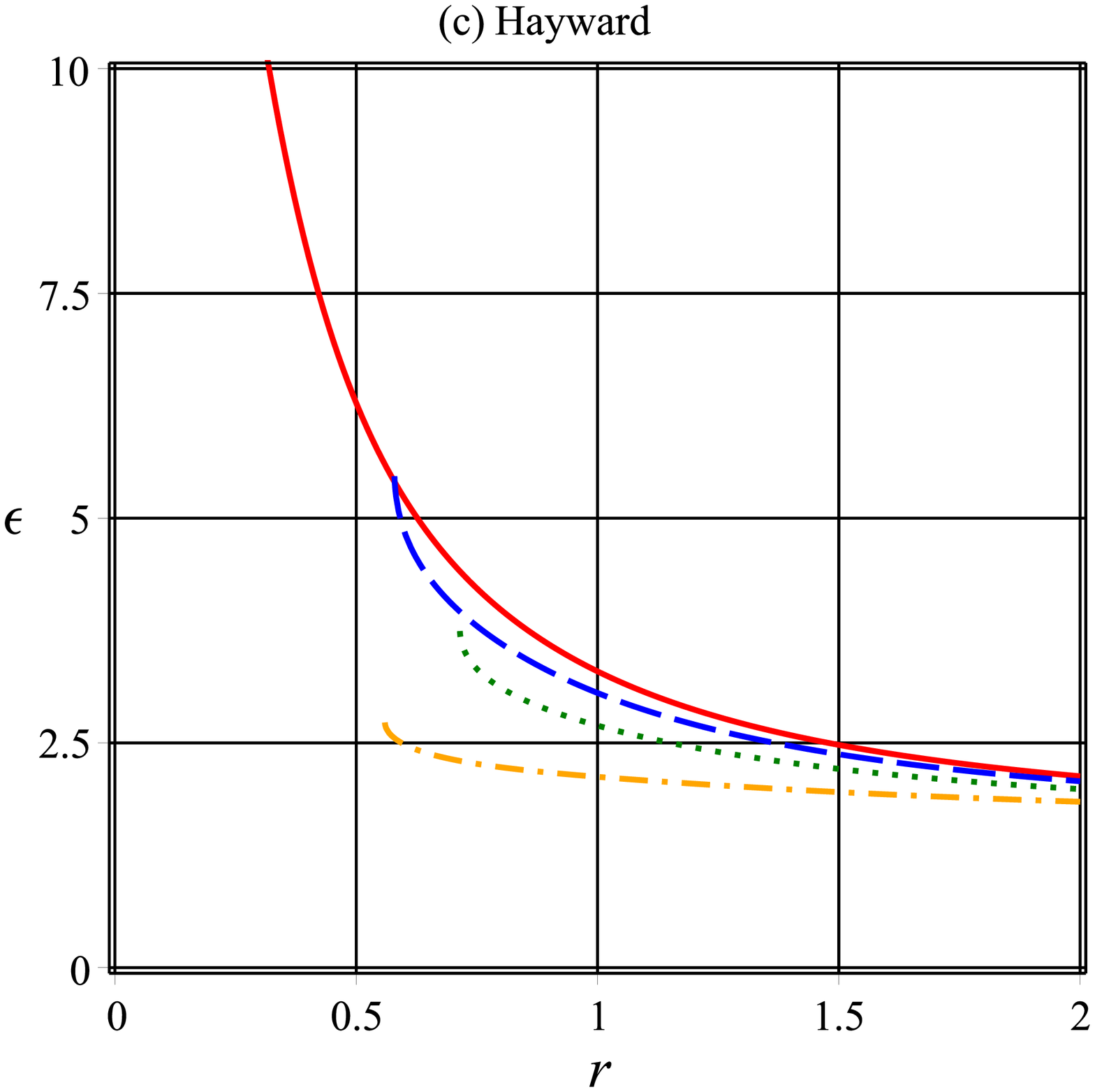}\\
\includegraphics[width=50 mm]{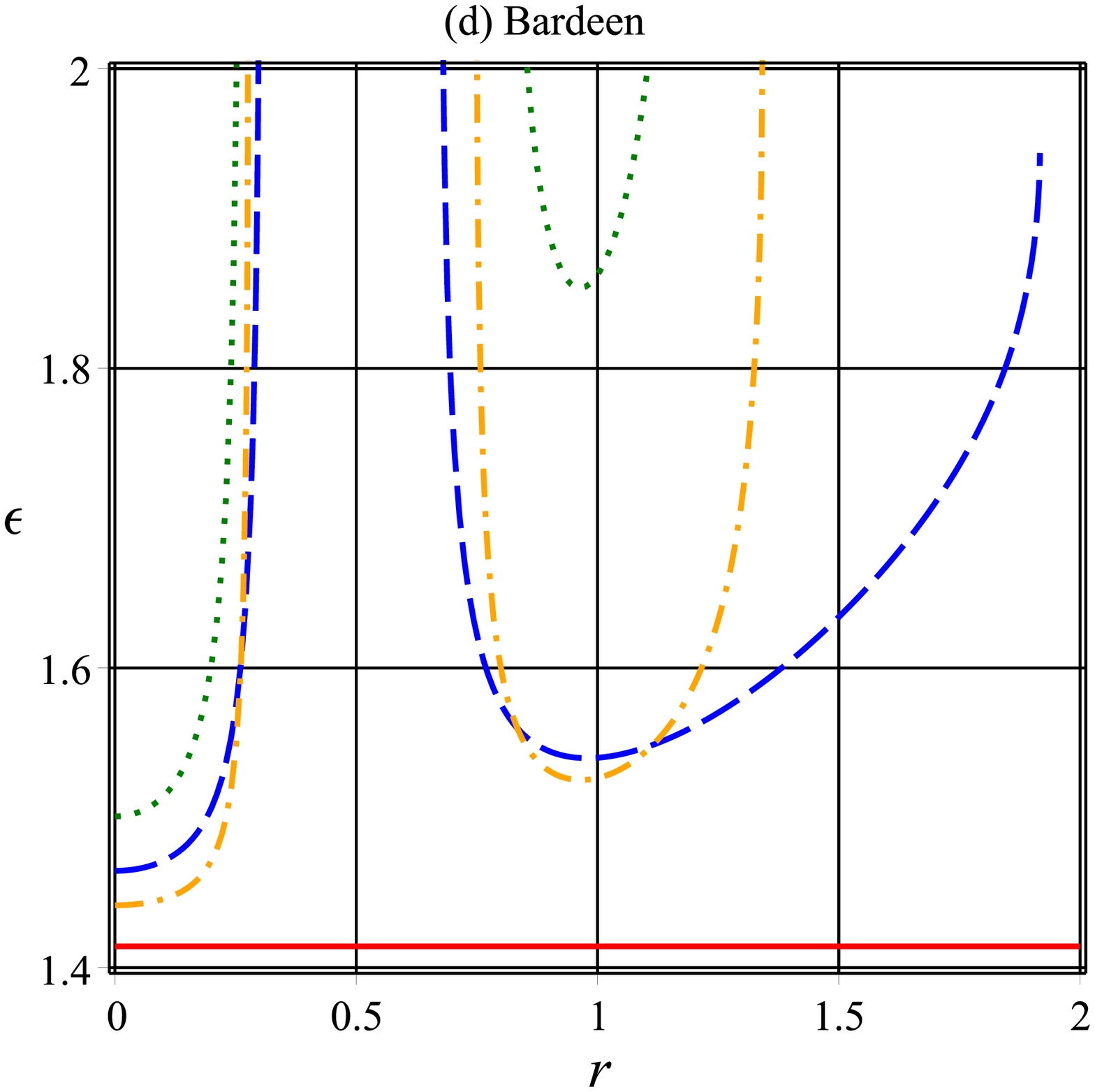}&\includegraphics[width=50 mm]{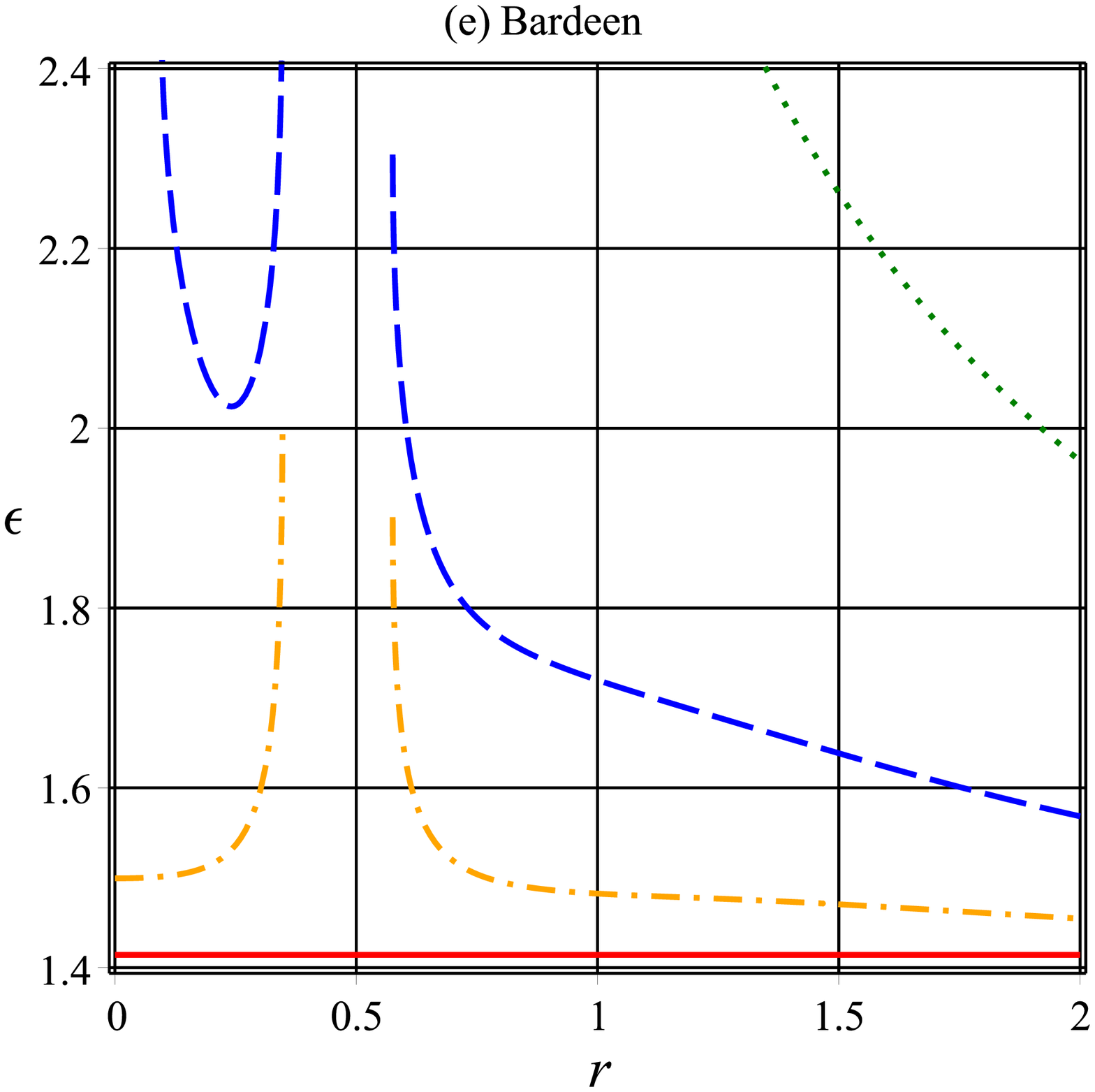}&\includegraphics[width=50 mm]{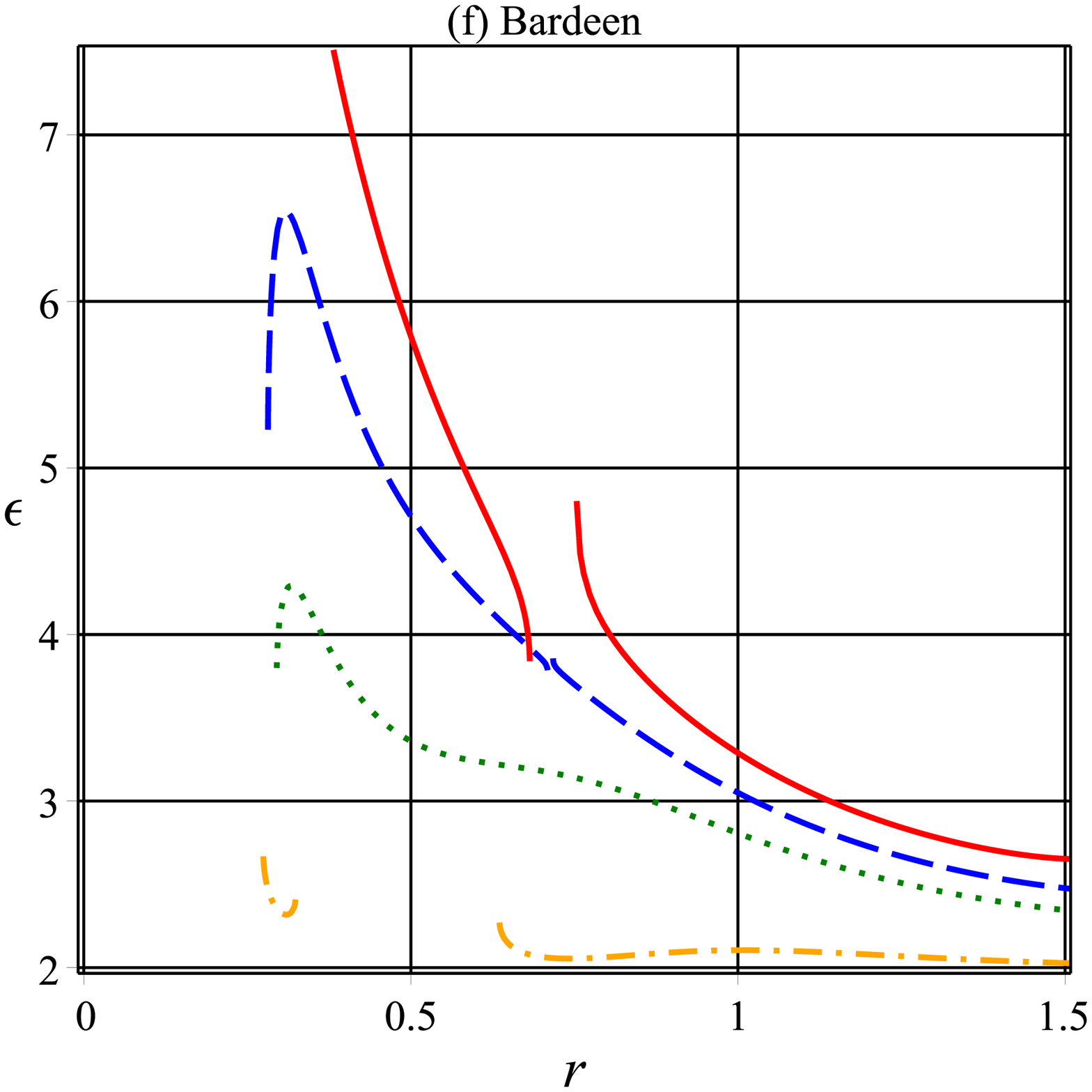}
 \end{array}$
 \end{center}
\caption{$\epsilon\equiv\tilde{E}_{CM}$ in terms of $r$ for
$\alpha=1$ and $\beta=2$  with $M=1$, $l=0.5$, $\mu=\nu=1$ and
$\theta=\frac{\pi}{2}$. (a) $a=0.5$, $L_{1}=L_{2}=2$.
$E_{1}=E_{2}=1$ (red solid), $E_{1}=2, E_{2}=1$ (blue dash),
$E_{1}=0, E_{2}=1$ (green dot), $E_{1}=0.4, E_{2}=1$ (orange dash
dot). (b) $a=0.5$, $E_{1}=E_{2}=2$. $L_{1}=L_{2}=2$ (red solid),
$L_{1}=0, L_{2}=2$ (blue dash), $L_{1}=-2, L_{2}=2$ (green dot),
$L_{1}=1, L_{2}=2$ (orange dash dot). (c) $E_{1}=1, E_{2}=2$,
$L_{1}=2, L_{2}=-2$. $a=0$ (red solid), $a=0.2$ (blue dash),
$a=0.4$ (green dot), $a=1$ (orange dash dot). (d) $g=0.5$, $a=0.5$, $L_{1}=L_{2}=2$.
$E_{1}=E_{2}=1$ (red solid), $E_{1}=2, E_{2}=1$ (blue dash),
$E_{1}=0, E_{2}=1$ (green dot), $E_{1}=0.4, E_{2}=1$ (orange dash
dot). (e) $g=0.5$, $a=0.5$, $E_{1}=E_{2}=2$. $L_{1}=L_{2}=2$ (red solid),
$L_{1}=0, L_{2}=2$ (blue dash), $L_{1}=-2, L_{2}=2$ (green dot),
$L_{1}=1, L_{2}=2$ (orange dash dot). (f) $g=0.5$, $E_{1}=1, E_{2}=2$,
$L_{1}=2, L_{2}=-2$. $a=0$ (red solid), $a=0.2$ (blue dash),
$a=0.4$ (green dot), $a=1$ (orange dash dot).}
 \label{fig:2}
\end{figure}

In this section we consider motion of particles with the rest mass
$m_{0}$ falling from infinity in the background of a rotating
modified Hayward or Bardeen black hole. Hamilton-Jacobi equation
governs geodesic motion of rotating modified Hayward or Bardeen
black hole which can written as,
\begin{equation}\label{T1}
\frac{\partial S}{\partial \tau}=-\frac{1}{2}g^{\mu\nu}\frac{\partial S}
{\partial x^{\mu}}\frac{\partial S}{\partial x^{\nu}},
\end{equation}
where $\tau$ is an affine parameter along the geodesics, and $S$
is the Jacobi action, so one can consider the following ansatz
\cite{Ghosh,Amir},
\begin{equation}\label{T2}
S=\frac{1}{2}m_{0}^{2}\tau-Et+L\phi+S_{r}(r)+S_{\theta}(\theta),
\end{equation}
where $S_{r}(r)$ and $S_{\theta}(\theta)$ are functions of $r$ and
$\theta$ respectively. Since equatorial motion
($\theta=\frac{\pi}{2}$) is assumed, so $S_{\theta}(\theta)=C$ is
possible choice, where $C$ is arbitrary constant. Moreover,
$E=-P_{t}$ and $L=P_{\phi}$ are conserved energy and angular
momentum respectively. One can obtain the null geodesic in the
following forms:
\begin{equation}\label{T3}
\dot{t}=\frac{1}{\xi}\left[aL(1-\tilde{h}\tilde{f})-a^{2}E(\tilde{f}-2)+E\Sigma\right],
\end{equation}
and
\begin{equation}\label{T4}
\dot{\phi}=\frac{1}{\xi}\left[L\tilde{h}\tilde{f}-a(1-\tilde{h}\tilde{f})E\right]
\end{equation}
where
\begin{equation}\label{T5}
\xi\equiv a^{2}\left(\tilde{h}\tilde{f}^{2}(\tilde{h}-1)+1\right)+\tilde{h}\tilde{f}\Sigma.
\end{equation}
and $dot$ measures the derivative with respect to the parameter
$\tau$. Using the Hamilton-Jacobi equation (\ref{T1}) and relation
(\ref{T2}), one can obtain,
\begin{equation}\label{T6}
-m_{0}^{2}=g^{tt}E^{2}-2g^{t\phi}EL+g^{\phi\phi}L^{2}+g^{rr}R^{2}(r),
\end{equation}
where
\begin{equation}\label{T7}
R(r)=\frac{dS_{r}(r)}{dr},
\end{equation}
and
\begin{eqnarray}\label{T8}
g^{tt}&=&\frac{a^{2}(\tilde{f}-2)-\Sigma}{\xi},\nonumber\\
g^{rr}&=&\frac{a^{2}+\tilde{f}\Sigma}{\Sigma},\nonumber\\
g^{\phi\phi}&=&\frac{\tilde{h}\tilde{f}}{\xi},\nonumber\\
g^{t\phi}&=&\frac{a(1-\tilde{h}\tilde{f})}{\xi},
\end{eqnarray}
It is easy to find,
\begin{equation}\label{T9}
R(r)=\sqrt{\frac{\Sigma}{a^{2}+\tilde{f}\Sigma}\left[\frac{2ELa(1-\tilde{h}\tilde{f})
-E^{2}(a^{2}(\tilde{f}-2)-\Sigma)-L^{2}\tilde{h}\tilde{f}}{\xi}-m_{0}^{2}\right]}.
\end{equation}
So, we have,
\begin{equation}\label{T10}
\dot{r}=\frac{a^{2}+\tilde{f}\Sigma}{\Sigma}R(r).
\end{equation}
Therefore, we have all non-zero 4-velocity components given by
equations (\ref{T3}), (\ref{T4}), and (\ref{T10}). Hence we are
able to obtain center of mass (CM) energy of two neutral particles
collision near the rotating modified Hayward or Bardeen black
hole. We suppose that the two particles have the same rest mass
($m_{0}$) with the angular momentum per unit mass $L_1$, $L_2$ and
energy per unit mass $E_1$, $E_2$, respectively. In that case CM
energy is given by,
\begin{equation}\label{T11}
\epsilon\equiv\tilde{E}_{CM}=\sqrt{1-g_{\mu\nu}u_{1}^{\mu}u_{2}^{\nu}},
\end{equation}
where $u_{i}^{\mu}=(\dot{t}_{i}, \dot{r}_{i}, 0, \dot{\phi}_{i})$,
$i=1,2$ (since for equatorial plane $\theta=\pi/2$, so
$\dot{\theta}=0$) and $E_{CM}=\tilde{E}_{CM}\sqrt{2}m_{0}$. After
some calculations we can find,
\begin{equation}\label{T12}
\tilde{E}_{CM}^{2}=\frac{1}{\xi^{2}}\left(\xi^{2}+\mathcal{A}L_{1}L_{2}+
\mathcal{B}E_{1}E_{2}+\mathcal{C}(E_{1}L_{2}+E_{2}L_{1})-H_{1}H_{2}\right),
\end{equation}
where
\begin{eqnarray}\label{T13}
\mathcal{A}&=&a^{2}\tilde{f}\tilde{h}(\tilde{f}\tilde{h}+\tilde{f}^{2}\tilde{h}
-\tilde{f}^{2}\tilde{h}^{2}-1)-\tilde{f}^{2}\tilde{h}^{2}\Sigma,\nonumber\\
\mathcal{B}&=&\tilde{f}\tilde{h}[a^{2}(\tilde{f}-2)-\Sigma]^{2}+a^{2}
(1-\tilde{f}\tilde{h})[a^{2}(\tilde{f}-2)-\Sigma],\nonumber\\
\mathcal{C}&=&a(1-\tilde{f}\tilde{h})\left(a^{2}(1-\tilde{f}\tilde{h})
-\tilde{f}\tilde{h}[a^{2}(\tilde{f}-2)-\Sigma]\right),
\end{eqnarray}
and
\begin{equation}\label{T14}
H_{i}=\sqrt{\xi}\sqrt{\xi m_{0}^{2}-2a(1-\tilde{f}\tilde{h})E_{i}L_{i}+[a^{2}(\tilde{f}-2)
-\Sigma]E_{i}^{2}+\tilde{f}\tilde{h}L_{i}^{2}},\hspace{1cm} i=1,2.
\end{equation}
We give numerical analysis of $\tilde{E}_{CM}$ given by the
equation (\ref{T12}) for both cases of rotating modified Hayward
and Bardeen black holes. From plots of the Fig. \ref{fig:2}, we
can see that for some cases of rotating modified Hayward, the CM energy has infinite value. On
the other hand it has finite constant value. From Figs.
\ref{fig:2} (a) and (d) we can see that $E_{1}=0$ or $L_{1}=0$
yields to infinite CM energy at horizon which is expected result.
On the other hand, from Figs. \ref{fig:2} (a) and (b) we can see
that $E_{1}=E_{2}$ and $L_{1}=L_{2}$ give constant CM energy with
finite value. Fig. \ref{fig:2} (c) shows variation of CM energy in
terms of rotational parameter $a$. We can see that,
infinitesimal $a$ may give infinite CM energy.\\

Near horizon limit $r\rightarrow r_{+}$ tells that,
\begin{equation}\label{T15}
\xi|_{r\rightarrow r_{+}}=(\tilde{h}-1)(\tilde{h}\tilde{f}^{2}-a^{2}).
\end{equation}
Therefore, CM energy will be infinite if we have $\tilde{h}=1$ or
$\tilde{h}\tilde{f}^{2}=a^{2}$. From green dotted line of Fig.
\ref{fig:2} (a) we can see infinite CM energy near inner
($r_{-}\approx0.9$) and outer ($r_{-}\approx1.65$) horizon. Also,
in the case of Bardeen black hole we can see from  Fig.
\ref{fig:2} (d) and (e) that the CM energy will be infinite.

\section{Particles Orbits}

In order to specify the range of the particles angular momentum,
we should calculate the effective potential for describing the
motion of the test particles. In the equatorial plane
($\theta=\pi/2$), the radial equation of motion for the time-like
particles moving along the geodesic is described by
\begin{equation}\label{P0}
\frac{1}{2}~\dot{r}^{2}+V_{eff}=0
\end{equation}

\begin{figure}[h!]
 \begin{center}$
 \begin{array}{cccc}
\includegraphics[width=50 mm]{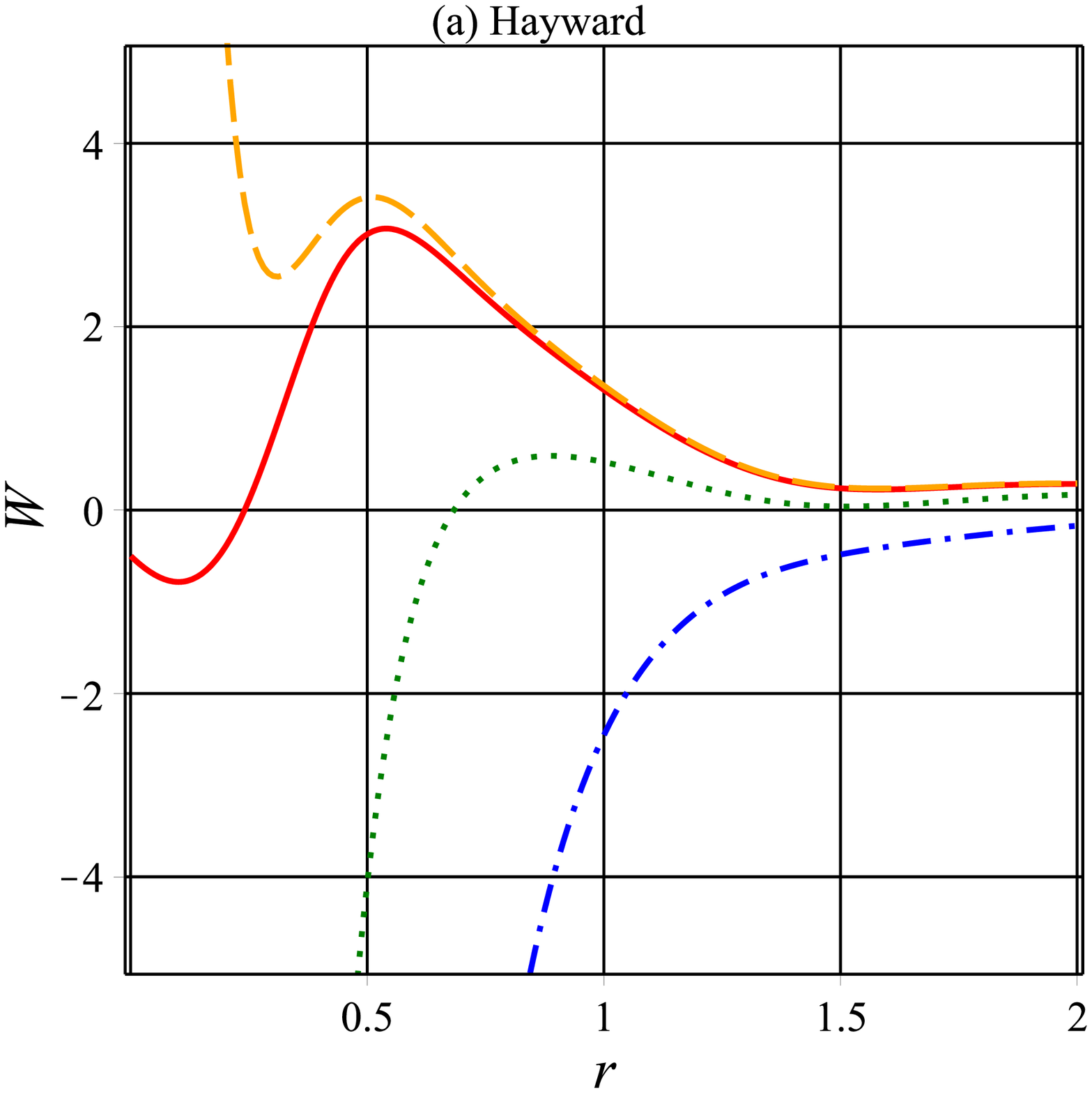}&\includegraphics[width=50 mm]
{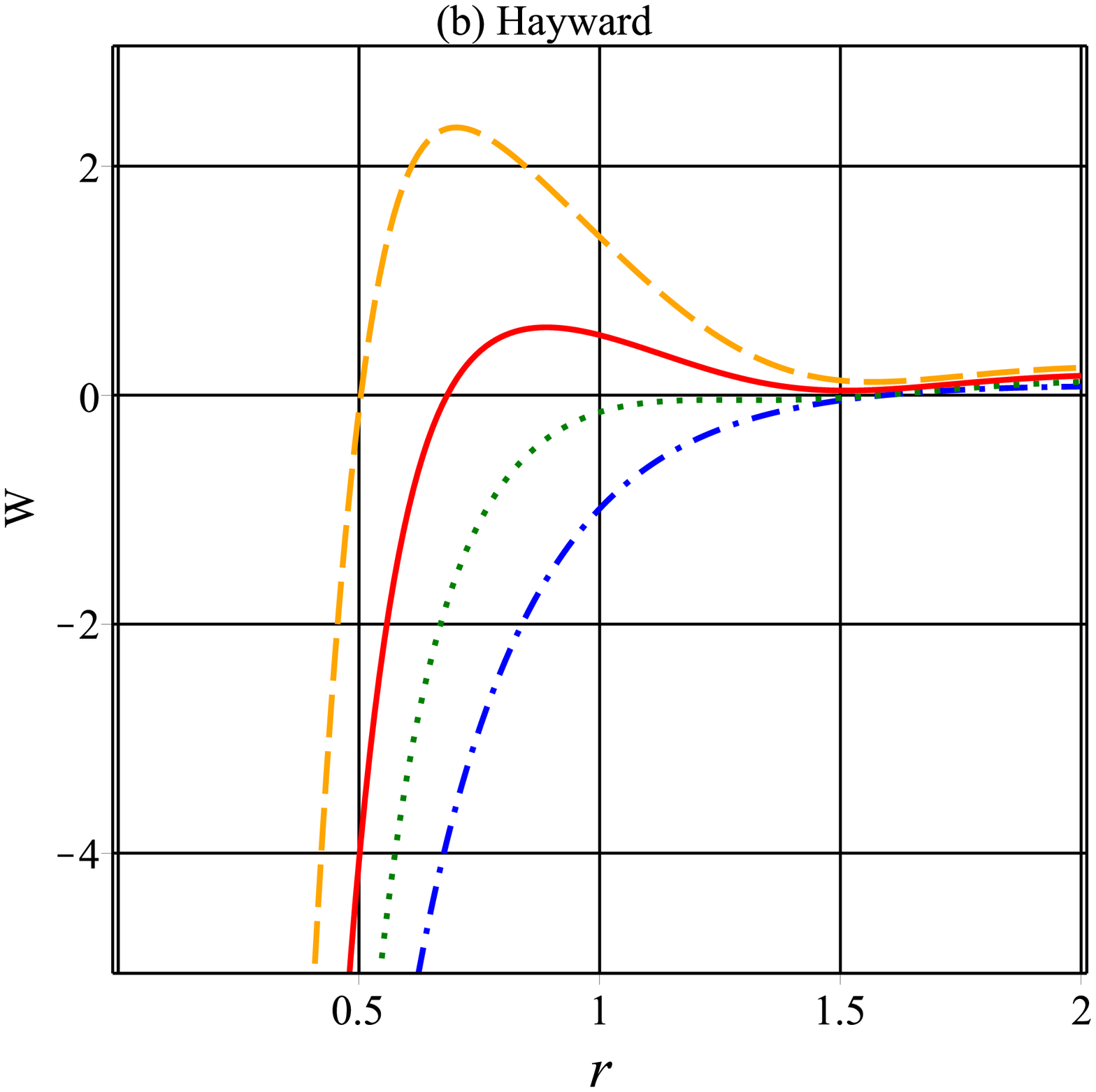}&\includegraphics[width=50 mm]{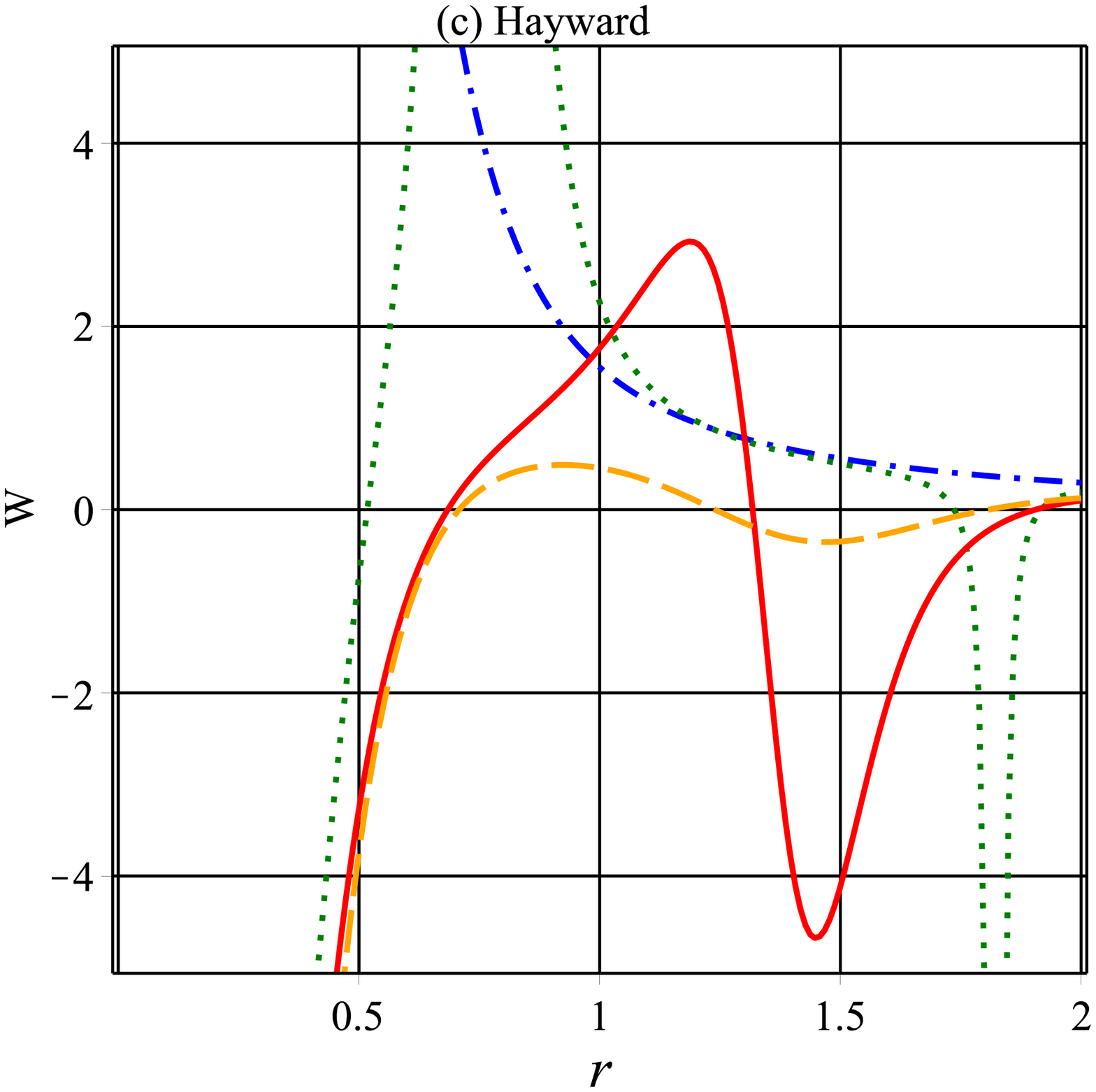}\\
\includegraphics[width=50 mm]{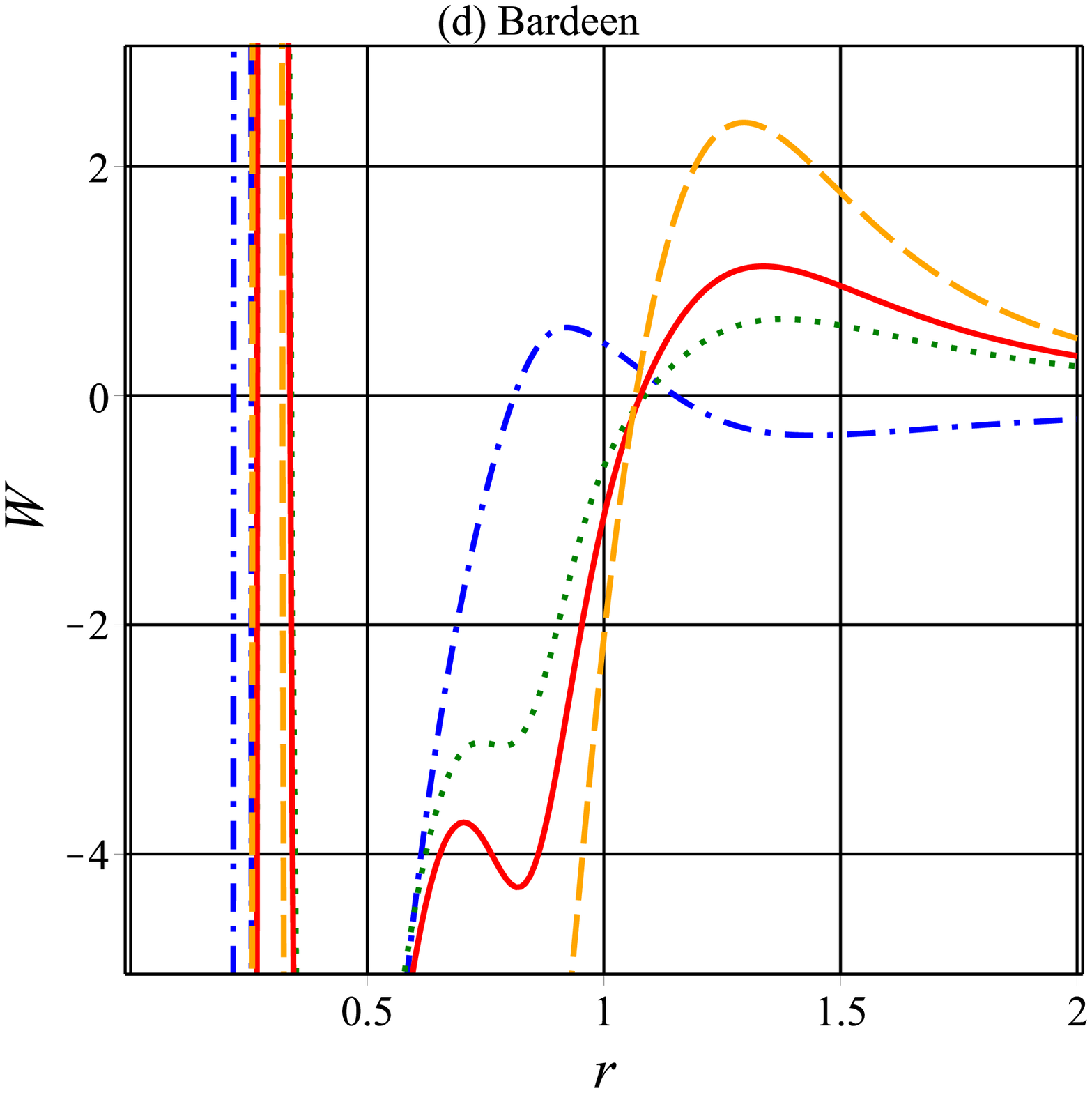}&\includegraphics[width=50 mm]
{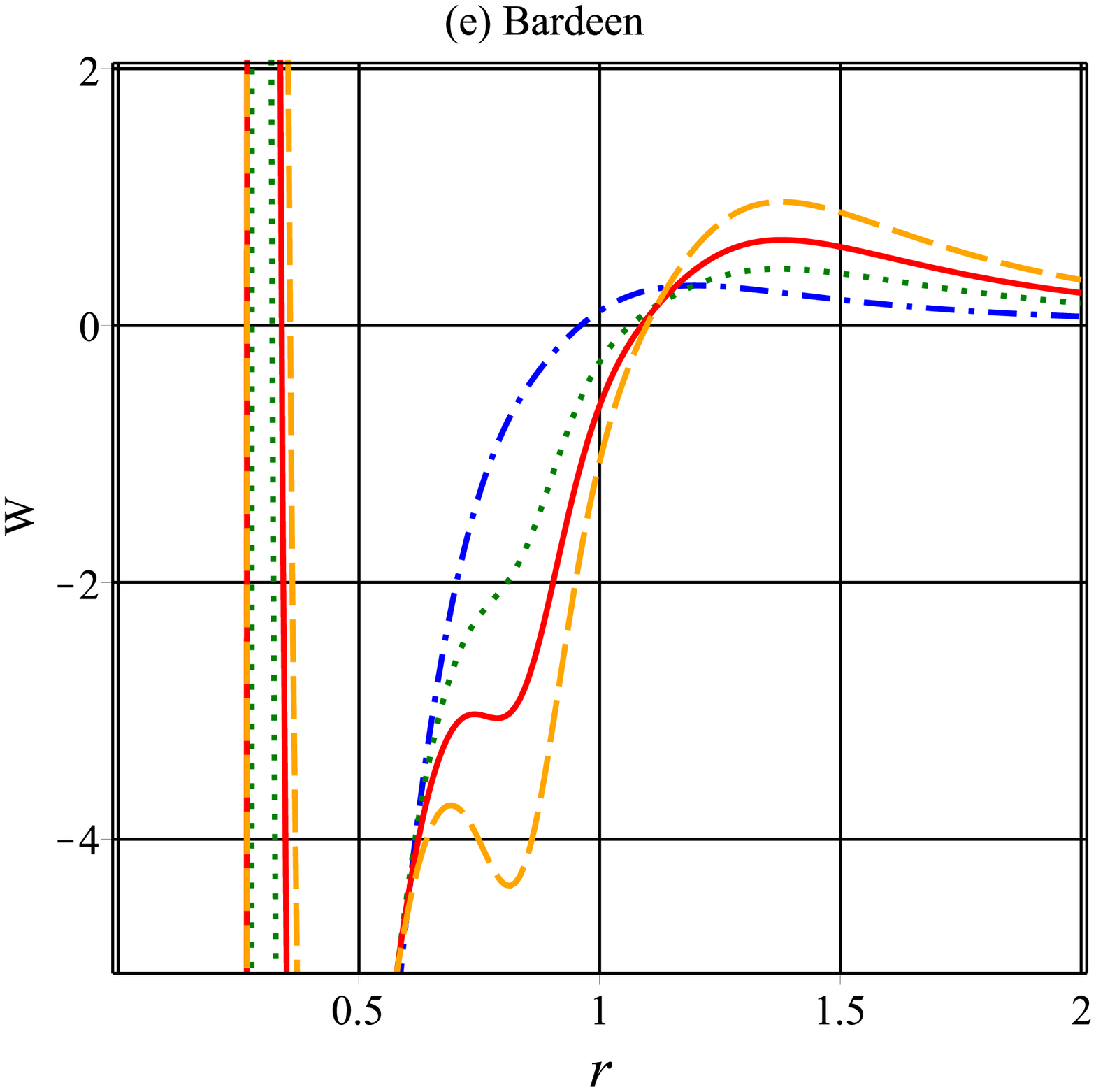}&\includegraphics[width=50 mm]{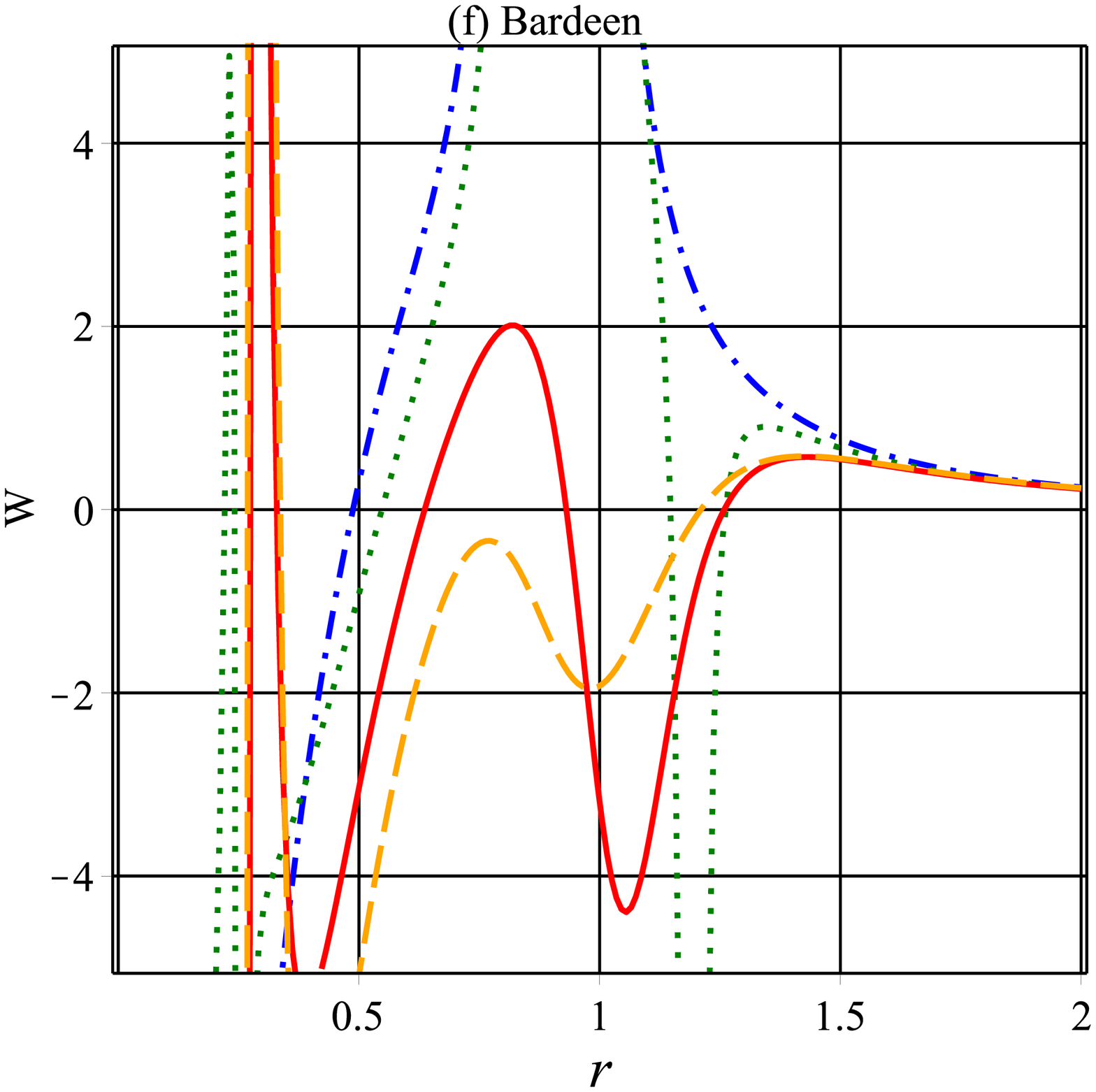}
 \end{array}$
 \end{center}
\caption{$W\equiv\frac{dV_{eff}}{dr}$ in terms of $r$ for
$\alpha=1$ and  $\beta=2$ with $M=1$, $l=0.5$, $g=0.25$ and
$\theta=\frac{\pi}{2}$. Upper plots drawn for Hayward black hole
and lower plots drawn for Bardeen black hole. (a) $a=1$, $E=1$,
$L=-2$ (blue dash dotted),  $L=-0.4$ (green dot), $L=0$ (red
solid), $L=0.02$ (orange dash). (b) $a=1$, $L=-0.4$, $E=0$ (blue
dash dotted),  $E=0.8$ (green dot), $E=1$ (red solid), $E=1.2$
(orange dash). (c) $E=1$, $L=-0.4$. $a=0$ (blue dash dotted),
$a=0.2$ (green dot), $a=0.6$ (red solid), $a=0.8$ (orange dash).
(d) $a=1$, $E=1$, $L=-2$ (blue dash dotted),  $L=-0.4$ (green
dot), $L=0$ (red solid), $L=0.8$ (orange dash). (e) $a=1$,
$L=-0.4$, $E=0$ (blue dash dotted),  $E=0.8$ (green dot), $E=1$
(red solid), $E=1.2$ (orange dash). (f) $E=1$, $L=-0.4$. $a=0$
(blue dash dotted),  $a=0.2$ (green dot), $a=0.6$ (red solid),
$a=0.8$ (orange dash).}
 \label{fig:3}
\end{figure}

which gives the effective potential,
\begin{eqnarray}\label{P1}
V_{eff}&=&-\frac{\dot{r}^{2}}{2}\nonumber\\
&=&-\frac{a^{2}+\tilde{f}\Sigma}{2\Sigma}\left[\frac{2ELa(1-\tilde{h}\tilde{f})
-E^{2}(a^{2}(\tilde{f}-2)-\Sigma)-L^{2}\tilde{h}\tilde{f}}{\xi}-m_{0}^{2}\right]
\end{eqnarray}
where we have used (\ref{T9}) and (\ref{T10}). The circular orbit
of the particles obtained using the following relations,
\begin{equation}\label{P2}
V_{eff}=0,
\end{equation}
and
\begin{equation}\label{P3}
W\equiv\frac{dV_{eff}}{dr}=0.
\end{equation}
The first condition (\ref{P2}) satisfied at black hole horizon. In
the Fig. \ref{fig:3} we can see the behavior of $W$ to satisfied
the condition. For the unitary values of $a$ and $E$ we can see
that condition (\ref{P3}) satisfied for the modified Hayward black
hole with negative $L$ (see Fig. \ref{fig:3} (a)). In the other
plots (Fig. \ref{fig:3} (b) and (c)) we can see effect of $a$ and
$E$. On the other hand Fig. \ref{fig:3} (d), (e) and (f) show that
rotating modified Bardeen black hole has no restriction for
negative $L$. For any values of $L$ we have particle circle.

\section{Extremal Limit}
\begin{figure}[h!]
 \begin{center}$
 \begin{array}{cccc}
\includegraphics[width=50 mm]{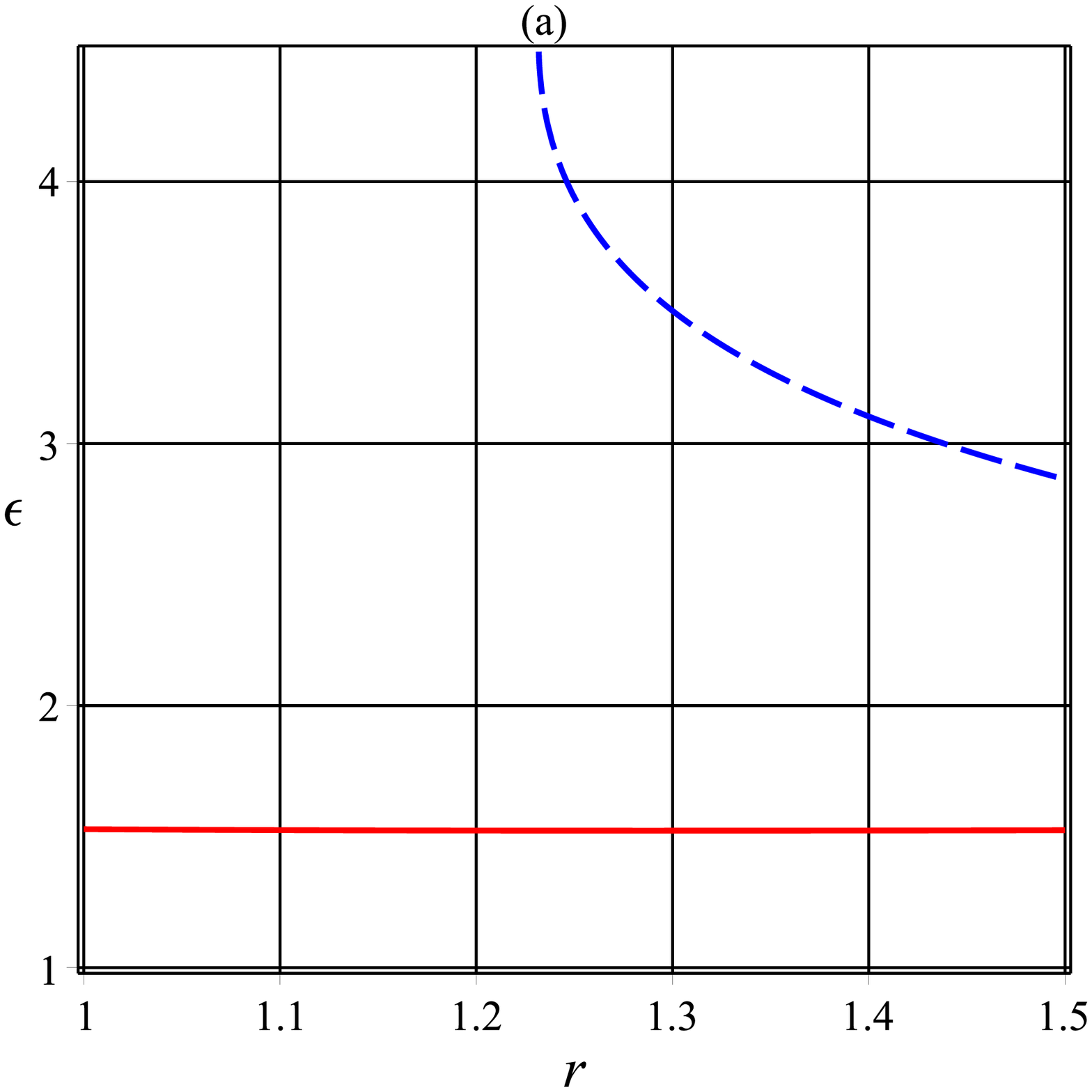}&\includegraphics[width=50 mm]{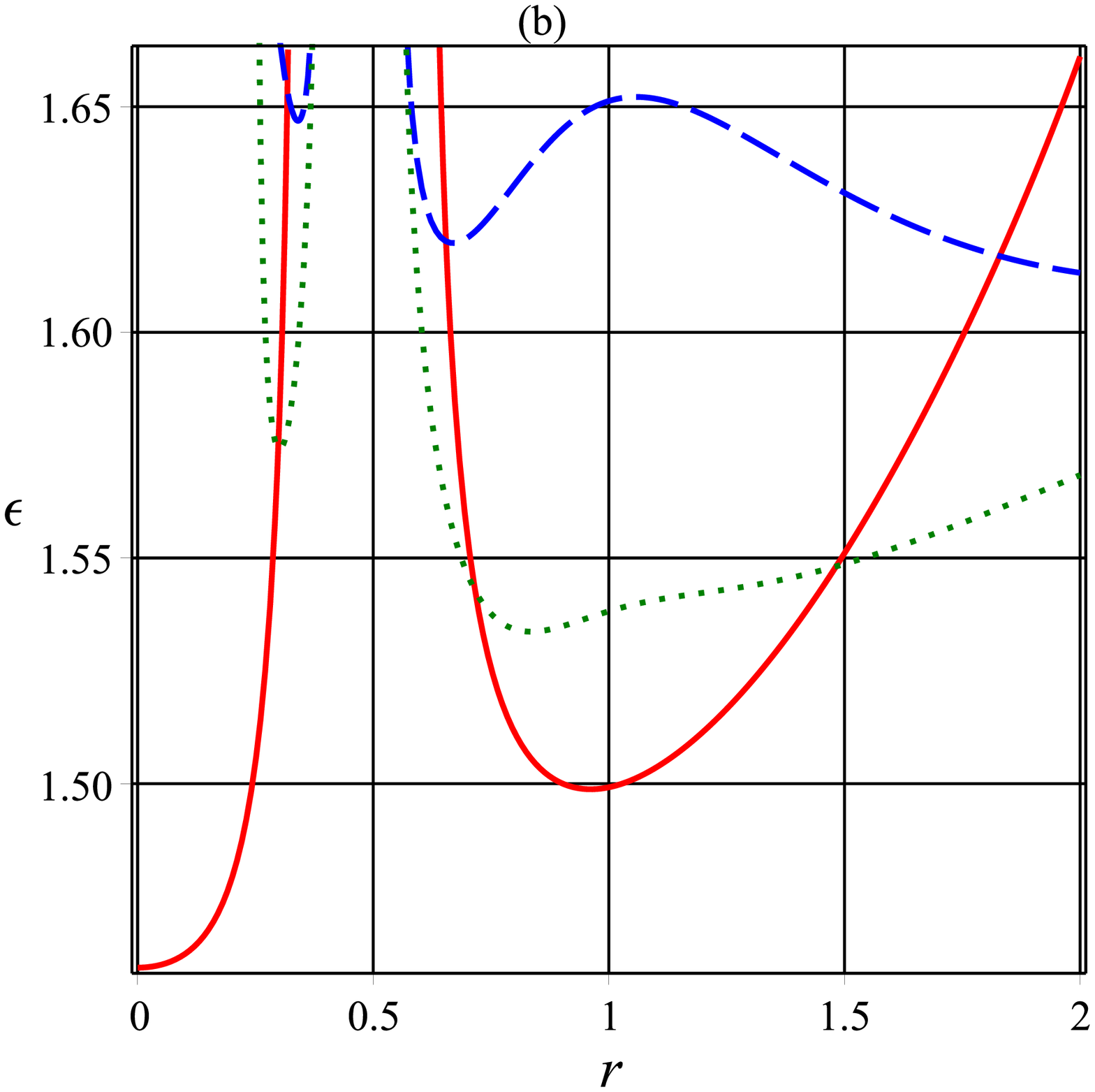}
 \end{array}$
 \end{center}
\caption{$\epsilon\equiv\tilde{E}_{CM}$ in terms of $r$ for
$\alpha=1$ and $\beta=2$ with $M=1$, $l=0.5$, $\mu=\nu=1$ and
$\theta=\frac{\pi}{2}$. (a) extremal ($a=0.65$) rotating modified Hayward black hole. $L_{1}=L_{2}=2$,
$E_{1}=1$, $E_{2}=2$ (red solid), $L_{1}=-2$, $L_{2}=2$, $E_{1}=1$, $E_{2}=2$ (blue dash),
$L_{2}=-2$, $L_{1}=2$, $E_{2}=1$, $E_{1}=2$ (blue dash).
(b) extremal ($a=1.22$) rotating modified Bardeen black hole. $L_{1}=L_{2}=2$,
$E_{1}=1$, $E_{2}=2$ (red solid), $L_{1}=0$, $L_{2}=2$, $E_{1}=1$, $E_{2}=2$ (blue dash),
$L_{2}=-2$, $L_{1}=0$, $E_{2}=1$, $E_{1}=2$ (green dot)}.
 \label{fig:4}
\end{figure}

It may be interesting to study solution at the extremal limit where $\delta=0$ and $r_{+}=r_{-}$. It will be obtained using appropriate choice of $\alpha$, $\beta$ and $a$. For example, extremal limit of rotating modified Hayward black hole may given by $\alpha=1$, $\beta=2$ and $a=0.65$ (other parameter fixed as previous), in that case $r_{+}=r_{-}\approx1.25$. Also extremal limit of rotating modified Bardeen black hole may given by $\alpha=1$, $\beta=2$, $g=0.5$ and $a=1.22$ (other parameter fixed as previous), in that case $r_{+}=r_{-}\approx0.6$. It is clear from Fig. \ref{fig:4} that infinite CM energy near the black hole horizon will be obtained for both Hayward and Bardeen black holes..

\section{Conclusions}
In this work, we have assumed two types of regular black holes
i.e., rotating modified Hayward and rotating modified Bardeen
black holes as particle accelerators. Horizon structure of
rotating modified Hayward black hole given by $g_{rr}=\infty$
which is exactly similar to the rotating Hayward black hole
discussed by the Ref \cite{Amir}. Horizon structure of rotating
modified Bardeen black hole given by the plots of Fig.
\ref{fig:1}. Figs. 1(a), 1(b) show $\Delta$ vs $r$ for
$\theta=\pi/2$ and Figs. 1(c), 1(d) show $\Delta$ vs $r$ for
$\theta=\pi/6$. We have investigated the the center of mass (CM)
energy of two colliding neutral particles with same rest masses
falling from rest at infinity to near the horizons of the
mentioned black holes. Figs. 2(a)-(f) show the CM energy vs $r$
for rotating modified Hayward and Bardeen black holes for
different cases. We have also investigated the range of the
particle's angular momentum and the orbit of the particle. Figs.
3(a)-(c) and Figs. 3(d)-(f) also show $W$ vs $r$ for rotating
modified Hayward and Bardeen black holes respectively. We have
also studied CM energy corresponding to extremal black holes and
obtained infinite CM energy for appropriate black hole
parameters and shown in the figures 4(a)-(b).\\

{\bf Acknowledgement:}\\

One of the author (UD) is thankful to IUCAA, Pune, India for warm
hospitality where part of the work was carried out.\\


\begin{thebibliography}{11}
\bibitem{BSW} M. Ba$\tilde{n}$ados, J. Silk and S. M. West, Phys. Rev. Lett. 103, 111102 (2009).
\bibitem{Lake1} K. Lake, Phys. Rev. Lett. 104, 211102 (2010).
\bibitem{Lake2} K. Lake, Phys. Rev. Lett. 104, 259903 (2010).
\bibitem{Harada} T. Harada and M. Kimura, Classical
Quantum Gravity 31, 243001 (2014).
\bibitem{Wei} S. W. Wei, Y. X. Liu, H. Guo and C.-E. Fu, Phys. Rev. D 82,
103005 (2010).
\bibitem{Liu} C. Liu, S. Chen, C. Ding and J. Jing, Phys. Lett. B 701, 285
(2011).
\bibitem{Zak} A. Zakria, M. Jamil, JHEP 1505 (2015) 147.
\bibitem{Berti} E. Berti, V. Cardoso, L. Gualtieri, F. Pretorius, U. Sperhake,
Phys. Rev. Lett. 103, 239001 (2009).
\bibitem{Jacob} T. Jacobson, T. P. Sotiriou, Phys. Rev. Lett. 104, 021101 (2010).
\bibitem{Zas} O. B. Zaslavskii, JETP Lett. 92, 571 (2010).
\bibitem{Igata} T. Igata, T. Harada and M. Kimura, Phys. Rev. D 85, 104028 (2012).
\bibitem{Ban} M. Banados, B. Hassanain, J. Silk and S. M. West, Phys. Rev. D
83, 023004 (2011).
\bibitem{Hus} I. Hussain, Mod. Phys. Lett. A 27, 1250017 (2012).
\bibitem{Grib} A. A. Grib, Yu.V. Pavlov, Astropart. Phys. 34, 581 (2011).
\bibitem{Har} T. Harada, M. Kimura, Phys. Rev. D 83, 024002 (2011).
\bibitem{Shar} M. Sharif and N. Haider, Astrophys. Space Sci. 346,
111 (2013).
\bibitem{Wei1} S. W. Wei, Y. X. Liu, H. T. Li and F.
W. Chen, JHEP 12, 066 (2010).
\bibitem{Ghosh} S. G. Ghosh, P. Sheoran, M. Amir, Phys. Rev. D 90 (2014) 103006.
\bibitem{Pour1} J. Sadeghi, B. Pourhassan, Eur. Phys. J. C.
72, 1984 (2012).
\bibitem{Pour2} J. Sadeghi, B. Pourhassan, H. Farahani, Commun. Theor. Phys. 62, 358 (2014).
\bibitem{P1} M. Patil and P. S. Joshi, Phys. Rev. D 82, 104049 (2010).
\bibitem{P2} M. Patil, P. S. Joshi and D. Malafarina, Phys. Rev. D 83, 064007
(2011).
\bibitem{P3}  M. Patil and P. S. Joshi, Classical Quantum
Gravity 28, 235012 (2011). \bibitem{P4}  M. Patil, P. S. Joshi, M.
Kimura and K. I. Nakao, Phys. Rev. D 86, 084023 (2012).
\bibitem{Bardeen} J. Bardeen, in Proceedings of GR5 (Tiflis, U.S.S.R., 1968).
\bibitem{Hayward} S. A. Hayward, Phys. Rev. Lett. 96, 031103 (2006).
\bibitem{Ayon}E. Ayon-Beato, A. Garcya , Phys. Rev. Lett. 80, 5056 (1998).
\bibitem{Abhas} G. Abbas · U. Sabiullah, Astrophys. Space Sci. 352, 769 (2014).
\bibitem{Bambi} C. Bambi and L. Modesto, Phys. Lett. B 721, 329 (2013).
\bibitem{Lor} T. De Lorenzo, C. Pacilioy, C. Rovelli and S.
Speziale, arXiv:1412.6015v1 [gr-qc].
\bibitem{Amir} M. Amir, S. G. Ghosh, arXiv:1503.08553 [gr-qc].
\bibitem{Pradhan} P. Pradhan, arXiv:1402.2748 [gr-qc].
\bibitem{Hai} N. Haider, Open J. Mod. Phys. 1, 1 (2014).
\bibitem{G} S. G. Ghosh and M. Amir, arXiv: 1506.04382 [gr-qc].
\end{thebibliography}
\end{document}